\documentclass{copernicus}

\begin{document}

%\linenumbers

\title{Cosmic rays as regulators of molecular cloud properties}

\author[1,2,*]{M. Padovani}
\author[3]{P. Hennebelle}
\author[2]{D. Galli}

\affil[1]{Laboratoire de Radioastronomie Millim\'etrique, UMR 8112 du CNRS, \'Ecole Normale Sup\'erieure et Observatoire de Paris, 24 rue Lhomond, 75231 Paris cedex 05, France}
\affil[2]{INAF--Osservatorio Astrofisico di Arcetri, Largo E. Fermi 5, 50125 
Firenze, Italy}
\affil[3]{CEA, IRFU, SAp, Centre de Saclay, 91191 Gif-Sur-Yvette, France}
\affil[*]{now at: Laboratoire Univers et Particules de Montpellier, UMR 5299 du CNRS,
Universit\'e de Montpellier 2, place E. Bataillon, cc072, 34095 Montpellier, France}

%% The [] brackets identify the author to the corresponding affiliation, 1, 2, 3, etc. should be inserted.

\runningtitle{Cosmic rays as regulators of molecular cloud properties}

\runningauthor{M. Padovani et al.}

\correspondence{M. Padovani\\ (Marco.Padovani@lupm.univ-montp2.fr)}

\received{}
\pubdiscuss{} %% only important for two-stage journals
\revised{}
\accepted{}
\published{}

%% These dates will be inserted by the Publication Production Office during the typesetting process.

\firstpage{1}

\maketitle  %% Please note that for the copernicus2.cls this command needs to be inserted after \abstract{TEXT}

\begin{abstract}
Cosmic rays are the main agents in controlling the chemical evolution
and setting the ambipolar diffusion time of a molecular cloud. We
summarise the processes causing the energy degradation of cosmic
rays due to their interaction with molecular hydrogen, focusing on
the magnetic effects that influence their propagation. Making use
of magnetic field configurations generated by numerical simulations,
we show that the increase of the field line density in the collapse
region results in a reduction of the cosmic-ray ionisation rate.
As a consequence the ionisation fraction decreases, facilitating
the decoupling between the gas and the magnetic field.
\end{abstract}

%% only used for copernicus2.cls
%\abstract{
% TEXT
% \keywords{TEXT}}

\introduction  %% \introduction[modified heading if necessary]

Low-mass prestellar cores are the basic units of star formation in
nearby clouds like Taurus and Perseus, where stars like our Sun
have been forming over the last few million years. The study of the
physical structure and kinematics of these cores is therefore crucial
for our understanding of the star formation process. Moreover,
prestellar cores are ideal laboratories for interstellar medium
chemistry which can be modelled using observations with different
tracers.

Cosmic rays (hereafter CRs) have a leading role in the dynamics and
chemistry of the interstellar medium (ISM).  The energy density of
CRs with energies $E\gtrsim1$~GeV is about 1~eV~cm$^{-3}$.  This
value is comparable to that present in the Galactic magnetic field,
to the energy density of the cosmic microwave background radiation,
and close to the local energy density in starlight~\citep{ww89,l02}.

%Some of this must be fortuitous, e.g. in globular clusters the energy density in starlight can be
%thousands of time the local value, but
%some may have an implication: the energy density in CRs must be related to magnetic fields, through their interaction, since
%CRs are charged particles.

We are witnessing an era of strong development of new telescopes
with higher and higher resolution allowing new observing techniques
so as to constrain the CR flux at energies lower than about 1~GeV.
Detections of OH$^{+}$ and H$_{2}$O$^{+}$ in low H$_{2}$ fraction
regions \citep{n10,g10}, enhanced CR ionisation rate (hereafter
\hbox{$\zeta^\mathrm{H_{2}}$}) in molecular clouds close to supernova remnants \citep{bb11,ch11},
observations of H$^{+}_{3}$ in diffuse clouds \citep{im12} and
towards the Galactic centre \citep{gu08} as well as $\gamma$
luminosity of molecular clouds \citep{m10} pose the question about
how to reconcile the high values of \hbox{$\zeta^\mathrm{H_{2}}$} estimated in diffuse
regions ($\zeta^\mathrm{H_{2}}\sim10^{-15}-10^{-16}$~s$^{-1}$) with those ones
measured in denser clouds ($\zeta^\mathrm{H_{2}}\sim10^{-17}-10^{-18}$~s$^{-1}$).

\section{The role of cosmic rays on physics and chemical composition
of molecular clouds}

\begin{figure}
\vspace*{2mm}
\begin{center}
\includegraphics[width=8.3cm]{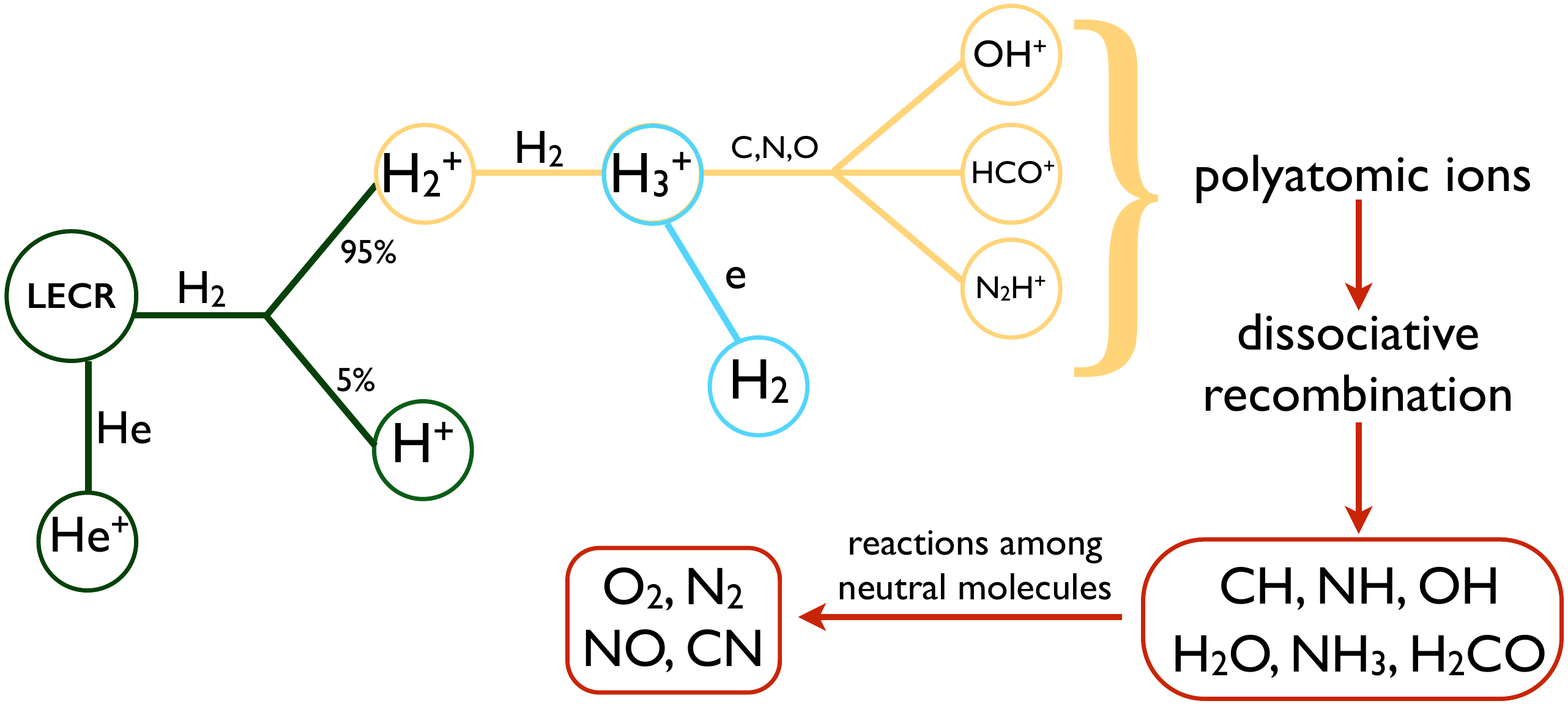}
\end{center}
\caption{The ionisation of molecular hydrogen due to an interaction
with a low-energy CR (LECR) leads to the formation of more and more
complex molecules that we can observe in molecular clouds. In
particular, in diffuse clouds, where the visual extinction $A_{V}$
is about 1~mag, the main reaction channel yields the trihydrogen
cation (H$_{3}^{+}$) that rapidly recombines with electrons. In
denser regions ($A_{V}>3-4$~mag) H$_{3}^{+}$ reacts with heavier
elements creating polyatomic ions up to neutral molecules, among
which ammonia and water.}
\label{chemistry}
\end{figure}

The study of the interaction of CRs with the ISM is a glaring example
of a multidisciplinary task involving the analysis of several
physical and chemical processes. In a prestellar core, the primary
source of ionisation is represented by CRs, since X-rays ionisation
arises only in presence of embedded young stellar objects~\citep{kk83,sn83}
and interstellar UV photons are absorbed in a thin layer of about
4 magnitudes of visual extinction \citep{mc89}.  The key quantity
that governs the interstellar chemistry, namely the creation of
more and more complex molecules in molecular clouds, is the so-called
{\em cosmic-ray ionisation rate}, that is the number of
hydrogen molecule ionisation per second (see e.g. \citeauthor{wh10}\
\citeyear{wh10}).  CRs interact with dense molecular clouds by
ionising their main component, the molecular hydrogen, and this
process activates the chemistry observed in clouds.  Since the
ionisation cross section of H$_{2}$ by collisions with electrons
and protons has a maximum at $\sim50$~eV and $\sim10$~keV~\citep{pg09},
respectively, the bulk of ionisation is due to low-energy CRs,
namely particles with energy lower than about $\mathrm{100~MeV-1~GeV}$
(see Fig.~\ref{chemistry}).

In turn, the ionisation fraction, that is the quantity of charged
particles with respect to neutrals that is proportional to $\sqrt{\hbox{$\zeta^\mathrm{H_{2}}$}}$
\citep{mo07}, controls the coupling of magnetic fields with the
gas, driving the dissipation of turbulence and angular momentum
transfer, thus playing a crucial role in protostellar collapse and
the dynamics of accretion discs \citep[e.g.][]{bh91,ph13}.

CRs also represent an important source of heating for molecular
clouds. In fact, inelastic collisions with interstellar molecules
and atoms convert about half of the energy of primary and secondary
electrons yielded by the ionisation process into heat
\citep[e.g.][]{gl73,gg12}.

During the last 50 years, several values of \hbox{$\zeta^\mathrm{H_{2}}$} ranging from a
few $10^{-16}$~s$^{-1}$ to a few $10^{-18}$~s$^{-1}$ have been
observationally determined in diffuse and dense interstellar clouds
from measurements of the abundances of various chemical species
\citep[see Fig.~6 in][and references therein]{pg13a}.  Nevertheless,
the lower limit of $\zeta^\mathrm{H_{2}}\sim10^{-17}$~s$^{-1}$ computed by \citet{st68}
is commonly used in chemical and magnetohydrodynamic (hereafter
MHD) as the ``standard'' \hbox{$\zeta^\mathrm{H_{2}}$} in molecular clouds.

Determining \hbox{$\zeta^\mathrm{H_{2}}$} from Earth is complicated because the interplanetary
magnetic field and the solar wind prevent low-energy CRs from
entering the heliosphere ({\em solar modulation}). This means that
Earth-based measurements of CR fluxes provide only a lower
limit on the interstellar spectrum of protons and heavy nuclei for
energies below $\sim1$~GeV/nucleon. Besides, the low-energy CR
electron flux, which already fluctuates in the energy range of
$10-100$~GeV \citep[see e.g.][]{cb04}, is damped by solar modulation.
This means that it is extremely difficult to know what happens to
the CR spectrum below GeV energies, representing the main constraint
for any trustworthy estimate of the CR ionisation rate in the ISM.
Only when the spacecrafts Voyager 1 and 2 will be far beyond the
heliopause, the outermost boundary for solar modulation effects
lying at $130-150$~AU from the Sun, it will be possible to measure
the low-energy CR spectrum.  On August 25$^\mathrm{th}$, 2012, Voyager
1 reached the heliopause and now we have information about the CR
interstellar spectrum down to energies of about 1~MeV \citep{pot13a}
and 10~MeV \citep{pot13b} for Galactic electrons and protons,
respectively, but still not enough to constrain the low-energy
cosmic-ray flux. Since Voyager 1 is escaping the solar system at a
speed of about 3.6~AU per year, in about 5 years from now we will
be able to look at the true interstellar spectrum.

\section{Energy losses and magnetic effects on cosmic-ray propagation}

\begin{figure} \vspace*{2mm} \begin{center}
\includegraphics[width=8.3cm]{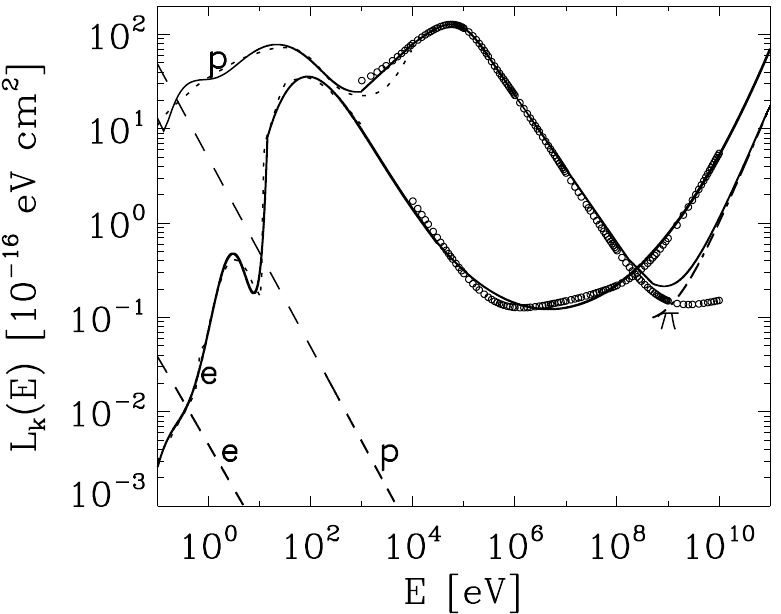} \end{center}
\caption{Energy loss functions $L_e(E_e)$ and $L_p(E_p)$ for electrons
and protons, respectively, colliding with H$_2$ ({\em solid} curves),
compared with NIST data ({\em circles}); {\em dashed} curves show
Coulomb losses for a fractional electron abundance $n_e/n(\mathrm{H}_2)=10^{-7}$; {\em dash-dotted} curve labelled with $\pi$ represents
the energy loss by pion production computed following \citet{s02};
{\em dotted} curves show the results by \citet{pav90} and \citet{dyl99}
for $p$--H$_2$ and $e$--H$_2$, respectively.} \label{elossfunctions}
\end{figure}

While crossing a molecular cloud, CRs undergo collisions with H$_{2}$
molecules. According to their initial energy and their composition,
they are slowed down due to processes that are specific of a
particular kind of particle (bremsstrahlung, synchrotron emission,
and inverse Compton scattering for electrons; elastic interactions,
pion production, and spallation for protons) or common both to CR
protons and electrons (Coulomb and inelastic interactions, and
ionisation). The {\em energy loss function} for the species $k$
is defined as
\begin{equation}
L_{k}(E_{k})=-\frac{1}{n(\mathrm{H_{2}})}\left(\frac{\mathrm{d} E_{k}}{\mathrm{d}\ell}\right)\,,
\end{equation}
where $n(\mathrm{H_{2}})$ is the density of the medium in which the
particle of energy $E_{k}$ propagates and $\ell$ is the path length.
Figure~\ref{elossfunctions} shows the energy loss functions for
protons and electrons colliding with molecular hydrogen.

While in the past it was assumed a lower cutoff for CR energy
to compute \hbox{$\zeta^\mathrm{H_{2}}$} \citep[e.g.][]{nb94}, in \cite{pg09} we showed
that even if a local interstellar spectrum is lacking of low-energy
particles, the slowing-down of CR protons and electrons during their
propagation produces a low-energy tail.  Our modelling is able to
explain the decrease of \hbox{$\zeta^\mathrm{H_{2}}$} with increasing hydrogen column density
computed from observations. In particular, a proton component at
low energies, and most likely also an electron component, could be
necessary to reproduce the data.

In \cite{pg11}, we studied how the presence of magnetic fields
affects the propagation of CRs. In fact, being charged particles,
CRs moves along field lines following an helicoidal path. This means
that they ``see'' a larger column density of molecular hydrogen
with respect to a rectilinear propagation, given by
\begin{equation}
N(\alpha)=\int_{0}^{\ell_\mathrm{max}(\alpha)}n(\ell)\,\mathrm{d}\ell\,,
\end{equation}
where $\ell_\mathrm{max}$ is the maximum depth reached inside the core
and $n(\ell)$ is the H$_{2}$ volume density. The angle $\alpha$,
called {\it pitch angle}, is the angle between the CR velocity and
the direction of the magnetic field and its evolution during the
CR propagation reads
\begin{equation}\label{alpha}
\alpha = \arccos\sqrt{1-\chi+\chi\cos^{2}\alpha_\mathrm{ICM}}\,,
\end{equation}
where $\chi=B/B_\mathrm{ICM}$ is the ratio between the local and the
intercloud magnetic field.  The two competing effects arising from
the presence of magnetic fields are {\em magnetic focusing} that
increases the CR flux where the field is more concentrated, and
{\em magnetic mirroring} according to which CRs are bounced out of
the cloud when the pitch angle reaches $\pi/2$, namely when the CR
velocity is perpendicular to the field line.

\section{Cosmic rays in collapsing clouds}

The decrease of \hbox{$\zeta^\mathrm{H_{2}}$} in the densest central regions of a protostellar
core may have a strong impact on the decoupling between gas and
magnetic field, leading the core towards the collapse.  \citet{ml09}
propose that the attenuation of \hbox{$\zeta^\mathrm{H_{2}}$} down to $10^{-18}$~s$^{-1}$
may increase the ambipolar diffusion having consequences on the
formation of a rotationally supported disc.

Following our previous studies \citep{pg09, pg11} where we
accurately examined the CR propagation accounting for column density
and magnetic effects, in \cite{ph13} we investigated the propagation
of CRs in the inner $300$--$400$~AU of a cloud core where the
formation of a protostellar disc is expected.  In particular, we
considered density and magnetic field configurations obtained by
ideal-MHD numerical simulations related to a rotating collapsing
core \citep{jh12}, performed with the AMR code RAMSES \citep{t02,fh06}.
One could deduce that magnetic effects are negligible when CRs reach
the inner part of a core (inside a radius of $\sim500$~AU). In fact
they may already be in the regime of exponential attenuation
($N>10^{25}$~cm$^{-2}$) since they have passed through a large
amount of column density \citep[see Fig.~1 in][]{ph13}.  On the
contrary, we found that even at very high densities magnetic fields
can efficiently remove CRs.

\begin{figure}
\vspace*{2mm}
\begin{center}
\includegraphics[width=8.3cm]{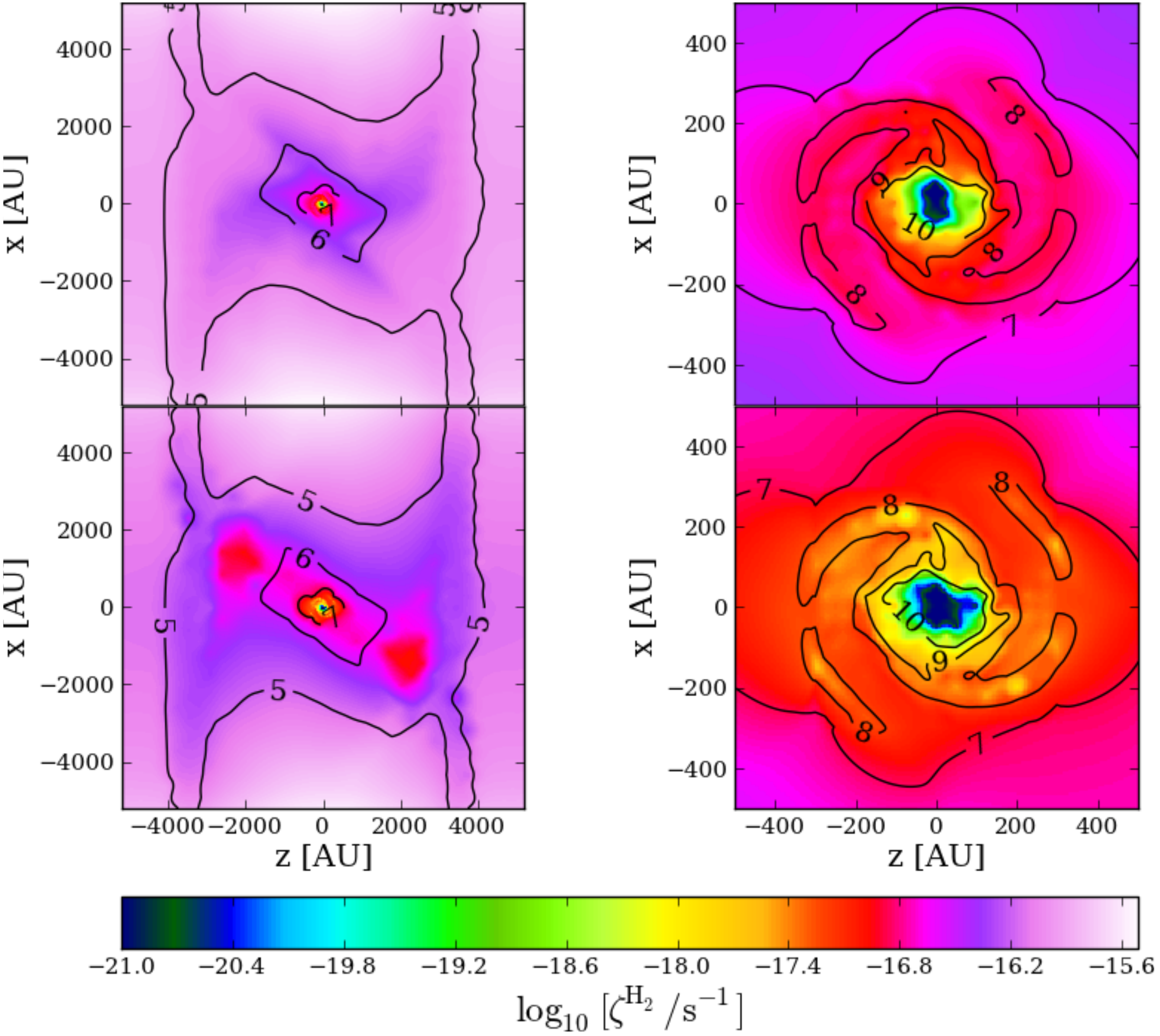}
\end{center}
\caption{CR ionisation maps and iso-density contours ({\em black
solid lines}) for the non-magnetic ({\em upper row}) and magnetic
({\em lower row}) cases. This model refers to a perpendicular
rotator, namely the main direction of the magnetic field and the
rotation axis are perpendicular, with a mass-to-flux ratio $\lambda=5$
\citep[see Fig.~10 in][for more details]{ph13}.  {\em Left} panels
show the entire computational domain while {\em right} panels show
a zoom in the inner region. Labels show $\log_{10}\ [n/\mathrm{cm^{-3}}]$.}
\label{zetamap-wandwoB}
\end{figure}

It is not possible to quantify to what extent column density
effects dominate over magnetic effects since it depends on the field
configuration considered. However, we did an estimate by calculating
\hbox{$\zeta^\mathrm{H_{2}}$} both accounting and neglecting magnetic effects.
Figure~\ref{zetamap-wandwoB} shows how magnetic shielding determines
a decrease of \hbox{$\zeta^\mathrm{H_{2}}$} by a factor of $\sim$10 at a radius of $300-400$~AU
and how the central region, where the minimum \hbox{$\zeta^\mathrm{H_{2}}$} is reached,
increases in size from $\sim$10 to $\sim$50~AU.  This example also
demonstrates that the use of the constant ``standard'' value
$\zeta^\mathrm{H_{2}}=10^{-17}$~s$^{-1}$ overestimates the CR ionisation rate in
the densest region of a molecular cloud.  Running our code for
different initial conditions (see Table~1 in \citeauthor{ph13}\
\citeyear{ph13}), we found a decrease of \hbox{$\zeta^\mathrm{H_{2}}$} below $10^{-18}$~s$^{-1}$
in the central $300-400$~AU, where $n\gtrsim10^{9}$~cm$^{-3}$, if
the toroidal component is larger than about 40\% of the total field
and in the cases of low and intermediate ionisation (mass-to-flux
ratio\footnote{It is a non-dimensional value that gives information
on the level of magnetisation.} $\lambda=17$ and 5, respectively).

In order to avoid running the whole code, we also formulated a
general fitting expression to approximately compute \hbox{$\zeta^\mathrm{H_{2}}$} as a
function of the column density, toroidal-to-poloidal magnetic field
ratio, and magnetic field strength (see Sect.~6 in \citeauthor{ph13}\
\citeyear{ph13}).

\conclusions  %% \conclusions[modified heading if necessary]

Cosmic rays constitute the main ionising and heating agent in dense,
starless, molecular cloud cores. We reexamined the physical quantities
necessary to determine the CR ionisation rate (especially the CR
spectrum below 1~GeV and the ionisation cross sections), and
calculated \hbox{$\zeta^\mathrm{H_{2}}$} as a function of the H$_{2}$ column density.  We
also accounted for magnetic effects, finding that mirroring and
focusing define the spatial domain where CRs can determine the
coupling between gas and magnetic fields.

Even if we are aware that the CR propagation should be computed
simultaneously with the MHD simulation, our study represents an
important proof of concept.  In fact, we showed that the inclusion
of magnetic effects is essential to account for the true path of
CRs during their propagation. We found that, in the densest region
of a protostellar core, \hbox{$\zeta^\mathrm{H_{2}}$} can be reduced of about $10^{3}$ times
the ``standard'' value of $10^{-17}$~s$^{-1}$, down to the lower
limit set by short-lived radionuclides in protoplanetary discs
\citep{un81,ca13}.

When the dynamical evolution becomes slower than the diffusion of
the magnetic field, the magnetic braking becomes inefficient. This
is predicted for densities larger that $10^{12}$~cm$^{-3}$ by
non-ideal MHD models \citep{db12}.  We noticed that in our models
the decrease of \hbox{$\zeta^\mathrm{H_{2}}$} can occur in some cases even at lower densities
($n>10^{9}$~cm$^{-3}$), resulting in very low ionisation fractions.
The consequences of the reduced CR ionisation rate on the magnetic
diffusion coefficients are analysed in detail in Padovani et
al.~(2014).

%\appendix
%\section{\\ \\ \hspace*{-7mm} HEADING}    %% Appendix A

%\subsection                               %% Appendix A1, A2, etc.

\begin{acknowledgements}
MP thanks K. Scherer, H. Fichtner, D. Bomans, K. Weis, and J. Tjus
for the invitation to this workshop.  MP and PH acknowledge the
financial support of the Agence National pour la Recherche (ANR)
through the COSMIS project.
\end{acknowledgements}


\begin{thebibliography}{}

\bibitem[Balbus \& Hawley(1991)]{bh91}
Balbus, S.~A. \& Hawley, J.~F. 1991, ApJ, 376, 214

\bibitem[Becker et al.(2011)]{bb11}
Becker, J.~K., Black, J.~H., Safarzadeh, M. \& Schuppan, F.
2011, ApJL, 739, L43

\bibitem[Casadei \& Bindi(2004)]{cb04}
Casadei, D. \& Bindi, V. 2004, ApJ, 612, 262 

\bibitem[Ceccarelli et al.(2011)]{ch11}
Ceccarelli, C., Hily-Blant, P., Montmerle, T., et al.
2011, ApJL, 740, L4

\bibitem[Cleeves et al.(2013)]{ca13}
Cleeves, L.~I., Adams, F.~C., Bergin, E.~A. \& Visser, R. 2013, ApJ, 777, 28

\bibitem[Dalgarno et al.(1999)]{dyl99} 
Dalgarno, A., Yan, M. \& Liu, W.\ 1999, ApJS, 125, 237 

\bibitem[Dapp \& Basu(2012)]{db12}
Dapp, W.~B., Basu, S. \& Kunz, M.~W.
2012, A\&A, 541, A35

\bibitem[Fromang \& Hennebelle(2006)]{fh06}
Fromang, S., Hennebelle, P. \& Teyssier, R. 2006, A\&A, 457, 371

\bibitem[Gerin et al.(2010)]{g10}
Gerin, M., De Luca, M., Black, J. et al.
2010, A\&A, 518, L110

\bibitem[Glassgold \& Langer(1973)]{gl73}
Glassgold, A.~E. \& Langer, W.~D.\ 1973, ApJ, 186, 859

\bibitem[Glassgold et al.(2012)]{gg12}
Glassgold, A.~E., Galli, D. \& Padovani, M. 2012, ApJ, 756, 157

\bibitem[Goto et al.(2008)]{gu08}
Goto, M., Usuda, T., Nagata, T., Geballe, T.~R., McCall, B.~J. et al.
2008, ApJ, 688, 306

\bibitem[Indriolo \& McCall(2012)]{im12}
Indriolo, N. \& McCall, B.~J. 2012, ApJ, 745, 91

\bibitem[Joos et al.(2012)]{jh12}
Joos, M., Hennebelle, P. \& Ciardi, A. 2012 A\&A, 543, 128

\bibitem[Krolik \& Kallman(1983)]{kk83} 
Krolik, J.~H. \& Kallman, T.~R.\ 1983, ApJ, 267, 610 

\bibitem[Longair(2002)]{l02}
Longair, M.~S.: High Energy Astrophysics, Cambridge University Press, 2002

\bibitem[McKee(1989)]{mc89} 
McKee, C.~F.\ 1989, ApJ, 345, 782 

\bibitem[McKee \& Ostriker(2007)]{mo07}
McKee, C.~F. \& Ostriker, E.~C. 2007,  ARA\&A, 45, 565

\bibitem[Mellon \& Li(2009)]{ml09}
Mellon, R.~R. \& Li, Z.-H. 2009, ApJ, 698, 922

\bibitem[Montmerle(2010)]{m10}
Montmerle, T., in High Energy Phenomena in Massive Stars, ASP Conf. Ser. 422, Mart\'i~J., Luque-Escamilla, P.~L. \& Combi, J.~A.
(Eds.); ASP: San Francisco, CA, p. 85, 2010

\bibitem[Nath \& Biermann(1994)]{nb94}
Nath, B.~B. \& Biermann, P.~L. 1994, MNRAS, 267, 447

\bibitem[Neufeld et al.(2010)]{n10}
Neufeld, D.~A., Goicoechea, J.~R., Sonnentrucker, P., et al.
2010, A\&A, 521, 10

\bibitem[Padovani et al.(2009)]{pg09}
Padovani, M., Galli, D. \& Glassgold, A.~E. 2009, A\&A, 501, 619

\bibitem[Padovani et al.(2011)]{pg11}
Padovani, M. \& Galli, D. 2011, A\&A, 530, A109

\bibitem[Padovani \& Galli(2013a)]{pg13a}
Padovani, M. \& Galli, D.: Cosmic-Ray Propagation in Molecular Clouds, in:
Cosmic Rays in Star-Forming Environments; Torres, D. F.,
Reimer, O. (Eds.); Advances in Solid State Physics; Springer: Berlin; Vol. 34;
pp. 61-82, 2013a

\bibitem[Padovani et al.(2013b)]{ph13}
Padovani, M., Hennebelle, P. \& Galli, D., A\&A 2013b, 560, A114

\bibitem[Padovani et al.(2014)]{pg14}
Padovani, M., Galli, D., Hennebelle, P., Commer\c con, B. \& Joos, M. 2014, A\&A submitted

\bibitem[Phelps(1990)]{pav90} 
Phelps, A.~V.\ 1990, J. Phys. Chem. Ref. Data, 19, 3 

\bibitem[Potgieter et al.(2013a)]{pot13a}
Potgieter, M.~S., Vos, E.~E., Nndanganeni, R.~R., Boezio, M. \& Munini, R.
2013a, {\tt http://arxiv.org/abs/1308.1666}

\bibitem[Potgieter(2013b)]{pot13b}
Potgieter, M.~S. 2013b, {\tt http://arxiv.org/abs/1310.6133}

\bibitem[Schlickeiser(2002)]{s02} 
Schlickeiser, R.: Cosmic Ray Astrophysics, Springer Berlin 2002

\bibitem[Silk \& Norman(1983)]{sn83}
Silk, J. \& Norman, C.\ 1983, ApJL, 272, L49 

\bibitem[Spitzer \& Tomasko(1968)]{st68} 
Spitzer, L.~J. \& Tomasko, M.~G. 1968, ApJ, 152, 971

\bibitem[Teyssier(2002)]{t02}
Teyssier, R. 2002, A\&A, 385, 337

\bibitem[Umebayashi \& Nakano(1981)]{un81}
Umebayashi, T. \& Nakano, T. 1981, PASJ, 33, 617

\bibitem[Wakelam et al.(2010)]{wh10}
Wakelam, V., Herbst, E., Le Bourlot, J., Hersant, F., Selsis, F. et al.
2010, A\&A 517, A21 

\bibitem[Wdowczyk \& Wolfendale(1989)]{ww89}
Wdowczyk, J. \& Wolfendale, A.~W.\ 1989,
ARNPS, 39, 43

\end{thebibliography}
\end{document}